\documentclass[aps,prb,twocolumn,notitlepage]{revtex4-1}
\pdfoutput=1
\usepackage{graphicx,amsmath,amssymb,amsfonts,dsfont,natbib,physics}
\usepackage{multirow}
\usepackage{lipsum,color}
\usepackage{verbatim}
\usepackage{commath}
\usepackage{bm}
\usepackage{float}
\usepackage{hyperref}
\usepackage{pifont}
\begin{document}

\title{The ``not-$A$", RSPT and Potts phases in an $S_3$-invariant chain}

\author{Edward O'Brien}
\affiliation{Rudolf Peierls Centre for Theoretical Physics, Clarendon Laboratory, Parks Road, Oxford, OX1 3PU, United Kingdom}
\author{Eric Vernier}
\affiliation{Rudolf Peierls Centre for Theoretical Physics, Clarendon Laboratory, Parks Road, Oxford, OX1 3PU, United Kingdom}
\author{Paul Fendley}
\affiliation{Rudolf Peierls Centre for Theoretical Physics, Clarendon Laboratory, Parks Road, Oxford, OX1 3PU, United Kingdom}
\affiliation{All Souls College, Oxford, OX1 4AL, United Kingdom}

\vskip1in
\date{\today}

\begin{abstract} 
We analyze in depth an $S_3$-invariant nearest-neighbor quantum chain in the region of a $U(1)$-invariant self-dual multicritical point. We find four distinct proximate gapped phases. One has three-state Potts order, corresponding to topological order in a parafermionic formulation. Another has ``representation'' symmetry-protected topological (RSPT) order, while its dual exhibits an unusual  ``not-$A$" order, where the spins prefer to align in two of the three directions. Within each of the four phases, we find a frustration-free point with exact ground state(s).  The exact ground states in the not-$A$ phase are product states,  each an equal-amplitude sum over all states where one of the three spin states on each site is absent. Their dual, the RSPT ground state, is a matrix product state similar to that of Affleck-Kennedy-Lieb-Tasaki. A field-theory analysis shows that all transition lines are in the universality class of the critical three-state Potts model. They provide a lattice realization of a flow from a free-boson field theory to the Potts conformal field theory.  
\end{abstract}

\maketitle
\section{Introduction and phase diagram}

Recent years have seen an explosion of interest in ``exotic'' phases of quantum matter, to the point where many phases formerly viewed as exotic appear quite naturally. The venerable quantum Ising chain is now understood to provide a fundamental example of topological order in its Majorana fermion form \cite{Kitaev2001}.  The Haldane phase of spin-1 antiferromagnet chains \cite{Haldane1983A,Haldane1983B}, with its special point analyzed by Affleck, Kennedy, Lieb and Tasaki (AKLT) \cite{Affleck1987,Affleck1988}, now provides the archetype of symmetry-protected topological (SPT) order \cite{Gu2009,Pollmann2010,Chen2011,Pollmann2012}. The AKLT ground state is one of the canonical examples of a matrix product state \cite{Verstraete2008}.  Another major line of development is in analyzing systems invariant under a $\mathbb{Z}_n$ symmetry. These ``parafermion'' systems exhibit a rich variety of interesting behavior, in particular providing a possible avenue to universal topological quantum computation \cite{Alicea2015}.

The aim of this paper is to analyze in depth a model displaying much of this interesting physics.  We study a nearest-neighbor $S_3$-invariant quantum chain near a $U(1)$-invariant critical point \cite{Vernier2019}, and show that four distinct phases meet at this multicritical point. One phase has three-state Potts order, which in the parafermionic formulation is topological order. Another has what we call ``representation'' SPT (RSPT) order protected by the $S_3$ permutation symmetry. Such order is protected by the symmetry if the Hilbert space is not changed, but is destroyed if a spin$-1/2$ degree of freedom is coupled to the end of the chain. A third phase exhibits an unusual ``not-$A$'' order, where the ground states prefer to avoid one of the three spin states at each site. The fourth phase is a disordered one, dual to the Potts ordered phase.

Our model provides a unified way of studying interrelations between RSPT, topological, and ordered phases. One feature of our model is that the RSPT phase has nice behavior under Kramers--Wannier duality. Namely, RSPT order is dual to that of the not-$A$ order, and the duals of the exact RSPT state are product states. Another unifying feature is that all the phase transition lines are in the same universality class, that of the critical three-state Potts model. One is a direct transition between the ordered Potts phase and RSPT phase.

\begin{figure*}[ht]
\vspace{-1cm}
\includegraphics[width=.9\linewidth]{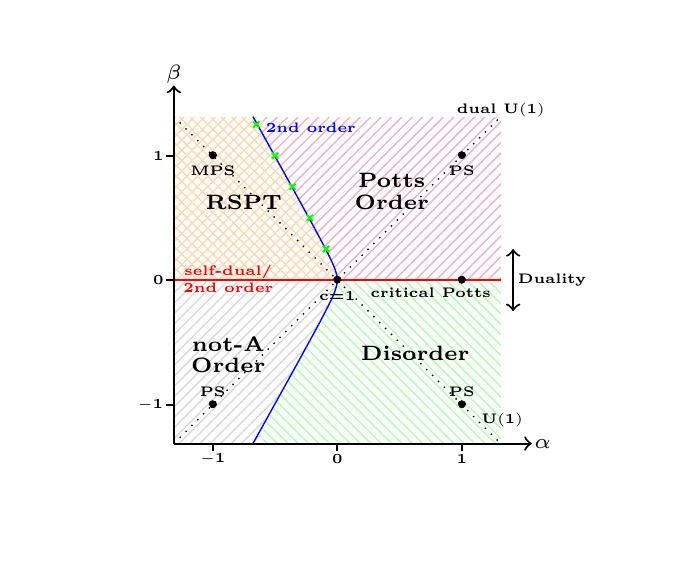}
\vspace{-2.7cm}\caption{The phase diagram surrounding the multicritical point, where the axes are defined via $J=\alpha+\beta$, $f=\alpha-\beta$ and $\lambda=1-\alpha$.  The multicritical point is at the center, with the points having (matrix) product ground states labelled by (M)PS. The duality $f\leftrightarrow J$ is a reflection about the line $\beta=0$. The horizontal line is self-dual (red), while the $U(1)$ and dual $U(1)$ (dotted black) symmetric lines are at 45 degree angles to it. ``Northeast" and ``northwest'' hatchings denote the ordered and disordered phases respectively, while crosshatching shows the RSPT phase.  The second-order transitions are along the solid lines, with the locations of the not-$A$/RSPT and Potts order-disorder transitions known exactly from duality. The other transitions are located numerically, as explained in section \ref{sec:numerics}, with the blue line an interpolation. 
}
\label{phase_diagram}
\end{figure*}

The Hamiltonian of our quantum chain (\eqref{H} below) is the linear combination of the Hamiltonian at the multicritical point with the usual three-state Potts Hamiltonian. We display our results for the two-parameter phase diagram in Figure \ref{phase_diagram}. The multicritical point is at the center, and four continuous phase transition lines meeting there separate the four phases.

A number of the features of this phase diagram can be seen in an effective field-theory description, simple enough to present here before doing the detailed analysis in section \ref{Continuum_limits}.  After the spatial averaging needed for the continuum limit, we find that the three-state system at each site of our quantum chain is described by a real bosonic field $\Phi(x)$. The three states correspond to $\Phi$ taking values $0$, $2\pi/3$ and $4\pi/3$, and we need to ``compactify'' the boson by identifying the values $\Phi\sim \Phi+2\pi$. The $\mathbb{Z}_3$ symmetry then corresponds to shifting $\Phi\to\Phi+2\pi/3$, while the charge-conjugation symmetry sends $\Phi\to-\Phi$.  The Euclidean (imaginary-time) action of our model valid near the multicritical point is written in terms of $\Phi$ and its dual field $\widehat\Phi$. It is
\begin{align}
\int d^2 x\left[ g(\nabla\Phi)^2 +  v\cos 3\Phi + \widehat{v}\cos 3\widehat{\Phi}\right]\,,
\label{SPhi}
\end{align}
where $g$, $v$ and $\widehat{v}$ are couplings. As the notation indicates, duality exchanges the latter two terms, while leaving the quadratic term invariant. Less obviously, we show in Section \ref{sec:freeboson} that duality requires $g=3/(4\pi)$. 

The multicritical point $v=\widehat{v}=0$ in \eqref{SPhi} is described by a free gapless bosonic field theory familiar in conformal field theory \cite{Ginsparg1988a,DiFrancesco1997} and in condensed-matter physics \cite{Chaikin1995,Giamarchi2003}. Here $\mathbb{Z}_3$ shift symmetry is promoted to a full $U(1)$ symmetry generated by $\Phi\to\Phi + a$ for any $0\le a<2\pi$.  Moreover, the self-duality requires dual $U(1)$ invariance under $\widehat{\Phi}\to\widehat{\Phi}+\widehat{a}$ for any $0\le \hat{a}<2\pi$. 

 Letting $v$ vary while keeping $\widehat{v}=0$ gives the gapped sine-Gordon field theory \cite{Mussardo2010}. The $S_3$ symmetry is spontaneously broken, because minima of the potential occur away from $\Phi=0$. These ordered phases can be characterized by the magnetization $M_g\equiv\bra{g}e^{i\Phi}\ket{g}$ in an $S_3$-breaking ground state $\ket{g}$. The symmetry requires $M_g^3$ to be real and independent of $g$.
 
The physics depends crucially on the sign of $v$.  For $v>0$, the values of $\Phi$ corresponding to lattice spin direction are minima of the potential. One expects, and our results confirm, conventionally ordered ground states with $M_g^3>0$. For $v<0$, these ground states are instead maxima of the potential, requiring the symmetry breaking to take a very different form, with $M_g^3<0$. We find that each of the three ground states prefers to exclude one of the three spin states at each site, while involving equal densities of the other two. Since we label the spin states as $A$, $B$ and $C$, we dub this phase the not-$A$ phase. 

By duality, the gap also occurs with $v=0$ and non-zero $\widehat{v}$.  The order parameter given by replacing $\Phi$ with $\widehat{\Phi}$ in $M_g$ must be non-vanishing. There can be no local order parameter, as along the $v=0$ line the full $U(1)$ symmetry is preserved by the Mermin--Wagner theorem. Indeed, $e^{\pm i\widehat{\Phi}}$ are non-local in the original field $\Phi$, as in the two-dimensional classical field theory, they create and annihilate vortex configurations in $\Phi$. 

Both phases at $\widehat{v}=0$ are disordered by the conventional definition. Nonetheless, they can be distinguished.  For $\widehat{v}>0$  vortices dominate, giving the Potts disordered phase.  For $\widehat{v}<0$ they are not as strongly favored, so different forms of long-range order are more likely. One of our central results is that this phase, the dual of the not-$A$ phase occurring at  $\widehat{v}<0$, has RSPT order. One way this form of long-range order manifests itself is by the presence of non-trivial degeneracies in the open chain. 

The action \eqref{SPhi} describes a gapped model in general. However, when $|v|=|\widehat{v}|$, it instead gives a field-theory description \cite{Lecheminant2002} of a flow from the free-boson field theory to the three-state Potts conformal field theory (CFT), a flow discovered by using exact scattering matrices and perturbed CFT \cite{Fateev1991,Delfino2002}. Our model therefore gives explicit lattice realizations of this flow. The self-dual case $v=\widehat{v}$ is the horizontal red line in Figure \ref{phase_diagram}. This Potts CFT thus describes not only the usual ferromagnetic Potts order/disorder transition, but the transition between not-$A$ order and the RSPT phase. Moreover, the field theory is independent of the signs of $v$ and $\widehat{v}$, so the same CFT must also describe the RSPT/Potts order and not-$A$/disorder transitions. These transition lines, however, cannot be exactly located in the lattice model, as it is not self-dual when $v=-\widehat{v}$.

The field-theory approach thus gives a good qualitative picture of the phase diagram. The purpose of this paper is to give a quantitative one by analysing the lattice model in detail. A key result is that at four points, one in each phase, we find exact ground states typifying the corresponding phase. An intriguing observation is that the eight exact ground states at these four points (three in each of the ordered phases, one in the disordered phase, and one in the RSPT phase), are precisely the eight possible conformal boundary conditions in the critical two-dimensional classical three-state Potts model \cite{Cardy1989,Affleck1998}. 

We start in section \ref{Model} by giving the Hamiltonian, and show that it is the most-general short-range Hamiltonian with the desired symmetries. The exact ground states are found in Section \ref{Exact_GS_section}, and the results extended to the full phases in Section \ref{sec:phases}, including a detailed analysis of the RSPT ground state and its relation to the AKLT state.
The field-theory action \eqref{SPhi} is derived in Section \ref{Continuum_limits}, with the arguments summarized here in the introduction explained in more depth.  Section \ref{sec:numerics} describes numerics completing the picture.

\section{The model}
\label{Model}
Throughout this paper we study a quantum spin chain with three states for each of the $L$ sites. The Hamiltonian and the symmetry operators acting on the $3^L$-dimensional Hilbert space are conveniently expressed in terms of the operators $\sigma_j$ and $\tau_j$ for $j=1,2,\dots L$, with $\sigma_{L+1}\equiv \sigma_1$. Each operator acts non-trivially on only one site $j$, e.g.\ $\sigma_j=1\otimes 1 \cdots 1\otimes \sigma\otimes 1\cdots 1$, so operators with different indices commute. We define
\begin{align}
\sigma_j^2=\sigma_j^{\dag}\,,\quad&\sigma_j^3=1\,,\quad \tau_j^2=\tau_j^{\dag}\,,\quad\tau_j^3=1\,,\cr &\sigma_j\tau_j=\omega\, \tau_j\sigma_j \ ,
\label{sigmataualg}
\end{align}
where $\omega=e^{2i\pi/3}$. 
In a basis where $\tau$ is diagonal the resulting matrix representation is
\begin{align}
\sigma = \begin{pmatrix}
0 & 1 & 0 \\
0 & 0 & 1 \\
1 & 0 & 0
\end{pmatrix}, \qquad \tau=\begin{pmatrix}
1 & 0 & 0 \\
0 & \omega & 0 \\
0 & 0 & \omega^2
\end{pmatrix}\ .
\label{tau_basis}
\end{align}

\paragraph*{Symmetries.}
We consider a two-parameter nearest-neighbour Hamiltonian invariant under the $S_3$ permutation group. 
This symmetry is generated by $\mathbb{Z}_3$ generator $\mathcal{S}$ and charge conjugation $\mathcal{C}$.  Their actions do not commute, and the fact that $S_3$ is non-abelian has interesting consequences.

Charge conjugation obeys $\mathcal{C}^2=1$. It acts on the operators as 
$\mathcal{C}\sigma_j\mathcal{C}=\sigma_j^{\dag}$ and 
$\mathcal{C}\tau_j\mathcal{C}=\tau_j^{\dag}$.
Cyclic permutations are generated by $\mathcal{S}=\prod_j \tau_j=\omega^Q$, where
\begin{align}
Q=\sum\limits_{j=1}^L S_j^z, \qquad S_j^z=\frac{i}{\sqrt{3}}\left(\tau_j^{\dag}-\tau_j\right)\ .
\label{U1_charge}
\end{align}
They obey $\mathcal{S}^\dagger\sigma_j\mathcal{S} =\omega \sigma_j$, while commuting with $\tau_j$. 
These permutations act diagonally in the basis \eqref{tau_basis}, while in the $\sigma$-diagonal basis they shift all spins. 
Since $\mathcal{S}^3=1$, the eigenvalues of  $\mathcal{S}$ are $\omega^s$, where $s=0,1,-1$. 

In some important special cases, the $\mathbb{Z}_3$ symmetry is enhanced to a full $U(1)$ symmetry generated by $Q$. 
Operators  having nice commutation relations with $Q$ are
\begin{align}
S_j^+=\frac{1}{3}\left(2-\omega\tau_j-\omega^{2}\tau_j^{\dag}\right)\sigma_j^{\dag} \ ,\qquad S_j^-=\left(S^+_j\right)^\dagger\ .
\label{Spmdef}
\end{align}
The corresponding single-site operators $S^\pm$ are proportional to the usual spin-1 $SU(2)$ raising and lowering operators and are given in the $\tau$-diagonal basis \eqref{tau_basis} by
\[ S^z=\begin{pmatrix}
0 & 0 & 0 \\
0 & 1 & 0 \\
0 & 0 & -1
\end{pmatrix},\ 
S^+=\begin{pmatrix}
0 & 0 & 1 \\
1 & 0 & 0 \\
0 & 0 & 0
\end{pmatrix},\ 
S^- = \begin{pmatrix}
0 & 1 & 0 \\
0 & 0 & 0 \\
1 & 0 & 0
\end{pmatrix}.
\]
They therefore satisfy
\begin{align}
\big[ S^z_j,\, S^{\pm}_j\big] =\pm S^\pm_j\ ,
\label{Scomm}
\end{align}
as of course can be derived directly from the algebra \eqref{sigmataualg}.

We impose two more discrete symmetries, as well as translation symmetry. One is parity, which simply exchanges the operators at $j$ and $L+1-j$: $\sigma_j\to\sigma_{L+1-j},\tau_j\to\tau_{L+1-j}$. The other is time reversal, which is anti-unitary and sends $\sigma_j\to\sigma_j^{\dag}$ but leaves $\tau_j$ invariant. This anti-unitary operation also sends any constant to its complex conjugate.

\paragraph*{Duality.} A very important property of our models is Kramers--Wannier duality, originally developed for the 2d classical Ising model \cite{Kramers1941}; see e.g.\ Ref.\ \onlinecite{Baxter1982} for its action in the Potts models. In our Hamiltonian setting with translation invariance and periodic boundary conditions, we can take duality to act on the operators as 
\begin{align}
\tau_{j}\to\sigma_j^{\dag}\sigma_{j+1}\ ,\qquad \sigma_j^{\dag}\sigma_{j+1}\to\tau_{j+1}\ .
\label{duality}
\end{align}
The charge $Q$ is therefore not invariant under duality, and so the dual of any Hamiltonian commuting with $Q$ must commute with the charge
\begin{align}
\widehat{Q} = \frac{i}{\sqrt{3}}\sum\limits\left(\sigma_j\sigma_{j+1}^{\dag}-\sigma_j^{\dag}\sigma_{j+1}\right)\ .
\label{dual_U1}
\end{align}

\paragraph*{The Hamiltonians.}

The three-state Potts model is the best-known $S_3$-invariant chain. With periodic boundary conditions, its Hamiltonian is
\begin{align}
H_P=-\sum\limits_{j=1}^L\left[ f\left(\tau_j+\tau_j^{\dag}\right) +J\left(\sigma_j^{\dag}\sigma_{j+1}+\sigma_j\sigma_{j+1}^{\dag}\right)\right]\ .
\label{H_P}
\end{align}
It is critical at the ferromagnetic and antiferromagnetic self-dual points, $J=f>0$ and $J=f<0$ respectively \cite{Baxter1982}.  The former separates an ordered phase $J>f\ge 0$ from the disordered phase $f>J\ge 0$. 

The other basic nearest-neighbour $S_3$ invariant Hamiltonian is much less known. It is
\begin{align}
H_0= \sum\limits_{j=1}^L \biggl[3\left(S_j^{{+}^2}S_{j+1}^{{-}^2}-S_j^+S_{j+1}^-+\text{h.c.}\right) -\tau_j-\tau_j^{\dag}\biggr]\ .
\label{H_c1}
\end{align}
This Hamiltonian obeys all the symmetries of $H_P$. In addition, $[Q,H_0]=0$, so the $\mathbb{Z}_3$ symmetry generated by $\mathcal{S}$ is enhanced to a $U(1)$ here. Moreover, $H_0$ is self-dual, as made apparent by the alternate form \eqref{H0TL} given below. It is a special case of the integrable spin-1 XXZ chain \cite{Alcaraz1989,Frahm1990}, and can be obtained from the anisotropic limit of the classical 19-vertex model \cite{Zamolodchikov1980}. Since it is both self-dual and commutes with $Q$, it must commute with $\widehat{Q}$ as well. It therefore possesses a large non-Abelian symmetry algebra generated by $Q$ and $\widehat{Q}$ known as the Onsager algebra. As detailed in our earlier work \cite{Vernier2019}, one of a number of interesting consequences is that the spectrum of $H_0$ has large degeneracies. 

In this paper we analyze the host of interesting physics coming from combining the two Hamiltonians via
\begin{align}
H(J,f,\lambda) =  \lambda H_0+ H_P(J,f) +(2\lambda-f-J)L\ .
\label{H}
\end{align}
We focus on the region around $H_0$, taking $\lambda\ge 0$.  In companion work \cite{OBrien2019b,OBrien2020}, we consider the equally interesting physics arising for $\lambda$ negative.

At $J=0$, our Hamiltonian has a $U(1)$ symmetry generated by $Q$, while at $f=\,$0, it has a $U(1)$ symmetry generated by $\widehat{Q}$.  In the phase diagram in Figure \ref{phase_diagram} we scale out an overall constant by parametrizing $J=\alpha+\beta$, $f=\alpha-\beta$ and $\lambda=1-\alpha$.  The point in the center of the diagram is $H_0$, where $f=J=\alpha=\beta=0$.   The self-dual line is the horizontal line, while the $U(1)$ lines are at 45 degree angles to it, with all meeting at $H_0$.

\paragraph*{Temperley--Lieb generators.} Expressing the Hamiltonian in terms of projection operators gives several useful insights. The basic Hermitian operators are defined as 
\begin{align}
p_{2j-1} &\equiv 1+\tau_j+\tau_{j}^\dagger\ ,\cr
p_{2j}&\equiv 1+\sigma^\dagger_j\sigma_{j+1}+\sigma_j\sigma_{j+1}^\dagger\ .
\label{pdef}
\end{align} 
They obey $p_a p_b=p_bp_a$ for $|a-b|>1$, and it is straightforward to show that
\begin{align}
(p_a)^2=3p_a\ ,\qquad p_{a}p_{a\pm 1} p_a = 3p_a\ .
\label{TLalg}
\end{align}
These relations, known as the Temperley--Lieb algebra \cite{Baxter1982}, are ubiquitous in the study of integrable lattice models and knot invariants. We have used an unconventional normalization (the usual being $e_a=p_a/\sqrt{3}$). The duality \eqref{duality} simply amounts to sending $p_j\to p_{j+1}$, consistent with the fact that (\ref{TLalg}) holds for all $a$, despite the different definitions for $a$ even and odd.

The Potts Hamiltonian obviously can be written as
\begin{align}
H_P = L(f+J) - \sum_{j=1}^L \left[ f\, p_{2j-1} + J\, p_{2j}\right]\ .
\label{HPTL}
\end{align}
Less obviously, 
\begin{align}
H_0 = -2L - \sum_{a=1}^{2L} \left(p_{a}p_{a+1} + p_{a+1} p_{a}+3p_{a}\right) \ .
\label{H0TL}
\end{align}
Although the latter expression obscures the $U(1)$ symmetry of $H_0$, it makes its self-duality apparent. The Hamiltonian \eqref{H0TL} for general Temperley--Lieb generators ($(e_a)^2=ne_a$ for general $n$) has been studied in depth in a different representation \cite{Ikhlef2009} with different physics.

\paragraph*{$H$ as a sum over projectors.}
The expressions (\ref{HPTL},\ref{H0TL}) are very useful in finding exact ground states at special points. 
The operator 
\begin{align}
P_{a,a+1}(\gamma) =  \gamma p_a + 3\gamma^{-1} p_{a+1} - p_a p_{a+1} - p_{a+1} p_a
\label{Pdef}
\end{align}
is proportional to a projector for any $\gamma$. Namely, the Temperley--Lieb algebra \eqref{TLalg} requires that
\begin{align}
\big(P_{a,a+1}(\gamma)\big)^2 = 3\big(\gamma  + 3\gamma^{-1}-2\big)P_{a,a+1}\ .
\label{Psq}
\end{align}
We then define
\begin{align}
H_{\gamma,\hat{\gamma}} \equiv \sum_{j=1}^{L} \left[P_{2j-1,2j}(\gamma) + P_{2j,2j+1}(\hat{\gamma})\right]\ .
\label{Hggdef}
\end{align}
Using the expression \eqref{H0TL} of $H_0$ in terms of the $p_a$ gives
\begin{align*}
H_{\gamma,\hat{\gamma}} 
=H_0 + 2L +\sum_{j=1}^{L}\Big[&\big(\gamma+3\hat{\gamma}^{-1}- 3\big)p_{2j-1} \cr
&+ \big(\hat{\gamma}+3{\gamma}^{-1} -3 \big)p_{2j}\Big]\ .
\end{align*}
With an appropriate definition of couplings, the last terms here are simply the Potts Hamiltonian \eqref{HPTL} up to a shift. Thus, fixing $\gamma$ and $\hat{\gamma}$ via
\begin{align} 
f= \lambda\big(3-\gamma -3\hat{\gamma}^{-1}\big)\, ,\quad J= \lambda\big(3-\hat{\gamma} - 3{\gamma}^{-1} \big)\, ,
\label{fJgamma}
\end{align}
our Hamiltonian \eqref{H} is rewritten as a sum over projectors as
\begin{align}
H(J,f,\lambda)= \lambda\,H_{\gamma,\hat{\gamma}}\ .
\label{HHgamma}
\end{align}
Requiring that the Hamiltonian be Hermitian means that either both $\gamma$ and $\hat\gamma$ are real, or $\gamma\hat{\gamma}^*=3$. 

When a hermitian Hamiltonian can be written as a sum over projectors with positive coefficients, all energies are non-negative. For the $\lambda$ positive case of interest here, the energies are thus non-negative when both $\gamma> 0$ and $\hat{\gamma}>0$, as is apparent from \eqref{Psq}. Moreover, any state annihilated by all the projectors is a ground state with energy zero.

\paragraph*{Other perturbations.}
Any other Hamiltonians preserving these symmetries are longer range in the following senses.
The only other nearest-neighbour terms obeying charge conjugation, time-reversal and parity will involve $p_{2j-1}p_{2j+1}$, whose dual is the next-nearest neighbor operator $p_{2j}p_{2j+2}$. 
Moreover, any such terms are longer range in the parafermion picture. Parafermions  \cite{Fradkin1980} are a generalisation of Majorana fermions useful for, among other things, understanding topological edge modes and their potential experimental applications \cite{Alicea2015}. Defining
\begin{align}
\psi_{2j-1}=\sigma_j\prod\limits_{k=1}^{j-1}\tau_k\ ,\quad \psi_{2j}=\omega\sigma_j\prod\limits_{k=1}^{j}\tau_k\ ,
\end{align}
we see that
\begin{align}
\tau_j=\omega^2\psi_{2j-1}^{\dag}\psi_{2j},\quad \sigma_j^{\dag}\sigma_{j+1}=\omega^2\psi_{2j}^{\dag}\psi_{2j+1}\ ,
\end{align}
so that the Potts chain is range two in terms of parafermionic operators, containing $\psi_a^\dagger\psi_{a+1}$ and its square.  The $U(1)$ invariant $H_0$ is range three, containing terms $\psi_a^\dagger\psi_{a+2}$, $\psi_a\psi_{a+1}\psi_{a+2}$ and their squares. Any other nearest-neighbor terms involve involve both $\tau_j$ and $\tau_{j+1}$ are range four, involving terms like $\psi_{a}^\dagger\psi_{a+3}$. Such terms are thus longer range when written in terms of the parafermions or the $p_a$. 

The $U(1)$-invariant $H_0$ is therefore the only other nearest-neighbour self-dual three-state Hamiltonian with all the symmetries of the Potts chain. This fact, along with the perturbed  conformal-field-theory arguments given in section \ref{Continuum_limits}, strongly suggest that all relevant operators obeying the desired symmetries are already included in the Hamiltonian defined in \eqref{H}.

\section{Exact ground states}
\label{Exact_GS_section}
To start our exploration of the phase diagram, in this section we analyze four special points where the exact ground states can be found. Happily, one such point occurs in each of the four phases, giving a great deal of insight into the types of ordering.

\subsection{Potts ordered and disordered points}
\label{sec:Pottsexact}

Setting $J=\lambda=0,\,f>0$ gives the Hamiltonian
\begin{align}
H_{\rm DP} = -fL-f\sum\limits_{j=1}^L \left(\tau_j+\tau_j^{\dag}\right).
\end{align}
The model is trivially solvable, with any state in the $\tau$-diagonal basis an eigenstate.
We denote the three eigenstates of $\tau_j$ by $\ket{0}$, $\ket{+}$ and $\ket{-}$, with eigenvalues $1,$ $\omega$ and $\omega^2$ respectively. The unique ground state is simply
\begin{align}
\ket{00...0}\ .
\end{align}
No local symmetry is spontaneously broken, and the model is gapped and completely disordered at this point.

Another trivially solvable point in the phase diagram occurs at the dual value  $f=\lambda=0,\,J>0$, where the Hamiltonian is simply 
\begin{align}
H_{\rm OP} = -JL-J\sum\limits_{j=1}^L \left(\sigma_j^{\dag}\sigma_{j+1}+\sigma_j\sigma_{j+1}^{\dag}\right)\ .
\label{Ordered_Potts}
\end{align}
Any state in the $\sigma$-diagonal basis is an eigenstate of $H_{\rm OP}$. We denote the three eigenstates of $\sigma_j$ on each site by $|A\rangle$, $\ket{B}$ and $|C\rangle$, with eigenvalues $1,$ $\omega$ and $\omega^2$ respectively. The three ground states of energy $-3L$ are then
\begin{align}
\ket{AAA\dots}\ ,\quad \ket{BBB\dots}\ ,\quad \ket{CCC\dots}\ .
\label{gsPotts}
\end{align}
The $\mathbb{Z}_3$ symmetry cyclically permutes these states, and so is spontaneously broken. Rewritten in terms of parafermions, this point is the simplest example of a $\mathbb{Z}_3$ topological phase \cite{Alicea2015}.

The ordered Potts point and the phase surrounding it are thus ordered and gapped. The most useful order parameter characterizing such a phase is the magnetization
\begin{align}
M_g = \bra{g}\sigma_j \ket{g}\ 
\label{Mdef}
\end{align}
in an $S_3$-breaking ground state $\ket{g}$. (We give an order parameter independent of ground state in \eqref{M3def} below.) As long as the ground state is translation invariant, $M_g$ will be independent of $j$. The three ground states \eqref{gsPotts} at the ordered point $H=H_{\rm OP}$ have
\begin{align}
M_A = 1\,,\quad M_B = \omega\,, \quad M_C =\omega^2\ .
\label{MPotts}
\end{align}
At the completely disordered point, the magnetization vanishes, as it must for any $S_3$-invariant ground state.

\subsection{Non-trivial exact ground states}
\label{sec:exactgs}

Two more special points possess exact ground states. These models are not trivially solvable like the Potts ordered and disordered points, and are not integrable. Knowing these ground states leads to a nice way of characterizing the phases on the left-hand side of the phase diagram in Figure \ref{phase_diagram}. 

We find these exact zero-energy ground states by utilizing the sum over projectors given in (\ref{Hggdef}, \ref{HHgamma}). 
Since the operators $P_{a,a+1}$ do not all commute with each other, zero-energy ground states do not occur for generic couplings. To find special points where they do, we note that having some of the projectors commute greatly simplifies the task of finding states annihilated by all the $P_{a,a+1}$. A little algebra shows that
\begin{align} 
\big[P_{a,a+1}(\gamma_0),\,P_{a+1,a+2}(\gamma_0') \big] = 0\ .
\end{align}
when \[ \gamma_0 = 3/2\ ,\quad \gamma_0'=2\ .\]
Both Hamiltonians $H_{3/2,2}$ and $H_{2,3/2}$  do indeed have exact ground states, as we show next.

\paragraph*{The ``not-$A$'' state.}
Setting $\gamma=\gamma_0$ and $\hat\gamma = \gamma_0'$ in the Hamiltonian translates to $f=0$ and $\lambda=-J>0$, as follows from \eqref{fJgamma}. Writing out the $P_{a,a+1}$ in the $\sigma$-diagonal basis allows one to see that $H_{3/2,2}$ has three exact ground states given by product states. Consider the nine states on the two sites $j,j+1$.  Then
\[
\ket{AA}-\ket{BA}-\ket{CA}\ ,
\]
is an eigenstate of  ${P}_{2j-1,2j}(\gamma_0)$ with non-zero eigenvalue (9/2 in our normalization). The two other states given by cyclic permutations under the action of the $\mathbb{Z}_3$ symmetry generators (i.e.\ all $A\to B \to C\to A$) are necessarily eigenstates with the same eigenvalue ${P}_{2j-1,2j}(\gamma_0)$. The other six states on the two sites are annihilated by $\mathcal{P}_{2j-1,2j}(\gamma_0)$. Likewise,
$P_{2j,2j+1}(\gamma_0')$ projects onto
\[
\ket{AA}-\ket{AB}-\ket{AC}
\]
and its cyclic permutations.  The state 
\begin{align}
\ket{BB}&+\ket{BC}+\ket{CB}+\ket{CC}\equiv 2\ket{\bar{A}}\otimes\ket{\bar{A}}\quad
\end{align}
is thus annihilated by $\mathcal{P}_{2j-1,2j}(\gamma_0)+P_{2j,2j+1}(\gamma_0')$, as are its 
cyclic permutations. We have denoted \[
\ket{\bar{A}}=\frac{1}{\sqrt{2}}\left(\ket{B}+\ket{C}\right)\ ,\]
with $\ket{\bar B}$ and $\ket{\bar C}$ given by cyclic permutations. 

Each term in the Hamiltonian $H_{3/2,2}$ thus annihilates the product states 
\begin{align}
\ket{\bar{A}\bar{A}...\bar{A}},\qquad \ket{\bar{B}\bar{B}...\bar{B}}, \qquad \ket{\bar{C}\bar{C}...\bar{C}}\ .
\label{gsnotA}
\end{align}
The ground state $\ket{\bar{A}\bar{A}\dots \bar{A}}$ is the equal-amplitude sum over all states not including $A$ on any site, and so we dub it the ``not-$A$'' state. Acting with the $\mathbb{Z}_3$ symmetry generators gives the not-$B$ and not-$C$ states, while charge conjugation exchanges not-$B$ and not-$C$. The $S_3$ symmetry is therefore spontaneously broken as in the Potts ordered phase.  The magnetizations are indeed non-vanishing, taking the values
\begin{align}
M_{\bar A} = -\frac{1}{2} \,,\quad M_{\bar B} = -\frac{1}{2}\omega\,, \quad M_{\bar C} =-\frac{1}{2}\omega^2\ .
\label{MnotA}
\end{align}

\paragraph*{The MPS state.}
An exact ground state also arises at the dual values $\hat{\gamma} = \gamma_0=3/2$ and ${\gamma}=\gamma_0'=2$, corresponding to $J=0$ and $\lambda=-f>0$.  Here the Hamiltonian is particularly simple, as using \eqref{H_c1} and \eqref{H} gives
\begin{align}
H_{2,\frac{3}{2}} = 3L + 3\sum\limits_{j=1}^L \left[S_j^{{+}}S_{j+1}^{{-}}\big(S_j^+S_{j+1}^- - 1\big) + \text{h.c.} \right].
\label{MPS_H}
\end{align}
As is manifest, this Hamiltonian is $U(1)$ invariant.  

The zero-energy ground state of $H_{2,3/2}$ is a matrix product state (MPS) of bond dimension 2. We group states at site $j$ into the matrix
\begin{align}
R(j)=\begin{pmatrix}
\ket{0} & \ket{+} \\
\ket{-}& \ket{0}
\end{pmatrix}\ .
\label{our_MPS}
\end{align}  
Using the form \eqref{MPS_H}, it is simple to check that
\[ \left(P_{2j-1,2j}(\gamma_0')+P_{2j,2j+1}(\gamma_0)\right)R(j)\otimes R(j+1) = 0\ ,
\] 
where the tensor product
\[
R(j)\otimes R(j+1) = \begin{pmatrix}
\ket{00}+\ket{+-} & \ket{0+}+\ket{+0} \\
\ket{-0}+\ket{0-} & \ket{-+}+\ket{00}
\end{pmatrix}
\]
is given by multiplying the two matrices. The full Hamiltonian therefore annihilates the state
\begin{align}
\ket{\psi_{\text{MPS}}} = \text{Tr}\left(R(1)\otimes R(2)\otimes \dots R(L)\right)\,,
\label{our_AKLT_GS}
\end{align}
where the trace is in the (suppressed) matrix indices (the ``auxiliary'' space), not in the Hilbert space. 

This zero-energy ground state $\ket{\psi_{\text{MPS}}}$ is invariant under the $S_3$ symmetry, and so has vanishing magnetization. It is straightforward to check that it is the unique ground state of $H_{2,3/2}$ for periodic boundary conditions. For open boundary conditions, however, there are four ground states, suggesting the existence of an SPT phase. We explain in section \ref{sec:phases} how it is not quite an SPT, but a less robust variation called a representation SPT.

\subsection{Connection to conformal boundary conditions}
We have found eight exact ground states at these four points, three in each of the ordered phases, and one in each of the other two.  We here point out an intriguing connection to the eight conformal boundary conditions in the three-state Potts conformal field theory (the very same CFT that describes the phase transitions in our model). Conformal boundary conditions do not introduce a length scale and so preserve (half of) the conformal symmetry. The boundary of a 2d model can be treated as a 1d state, and on the lattice, the space of all possible boundary conditions forms a vector space like the Hilbert space of our quantum chain. Powerful CFT tools \cite{Cardy1989} allow boundary states corresponding to conformal boundary conditions to be characterised and classified. 

For the three-state Potts CFT,  Refs.\ \onlinecite{Cardy1989,Affleck1998}, found the eight conformal boundary states, and gave intuition into them using of the two-dimensional classical Potts model at its critical point.  Remarkably, these eight states are precisely our ground states. The three states $\ket{AAA\dots}$,  $\ket{BBB\dots}$ and $\ket{CCC\dots}$, are the three fixed boundary conditions in the 2d model. The state $\ket{000\dots}$ is an equal-amplitude sum over all states in the $\sigma$-diagonal basis, and corresponds to free boundary conditions in the classical model. Three more states are called ``mixed''. In each mixed state, one of the three spin values is forbidden, just as in our not-$A$ states. The eighth conformal boundary state was uncovered in Ref.\ \onlinecite{Affleck1998}, and named ``new''. It proved more difficult to characterize on the lattice, but was shown to be the dual of the not-$A$ states. Thus it is precisely our MPS \eqref{our_AKLT_GS}!

\section{The phases}
\label{sec:phases}

In the preceding section \ref{Exact_GS_section}, we found four points with exact ground states, two ordered and two disordered. In this section we show that each is indicative of a distinct phase, and so our phase diagram must have at least four phase transition lines separating them. The distinct characteristics of each phase 
are summarized in Table \ref{Phase_table}.
\begin{table}[h]
\begin{tabular}{|c|c|c|c|}
\hline
Phase & $\ M^3\ $ &  Degeneracy  \\
\hline
Ordered Potts & $>0 $& 3 \\
\hline
Disordered Potts & 0 & 1 \\
\hline
RSPT & $0$ & 4 \\ 
\hline
not-$A$ & $<0$ & 3 \\
\hline
\end{tabular}
\caption{\label{Phase_table} The behavior of the order parameter $M^3$ and the ground-state degeneracy with open boundary conditions in the four gapped regions.}
\end{table}

\subsection{The ordered phases}

The two special points with ordered ground states both occur when $f=0$. The perfectly ordered Potts ground states \eqref{gsPotts} occur at $\lambda=0$, while three ground states of ``not-$A$'' type in \eqref{gsnotA} occur at $J=-\lambda$. All transform non-trivially under the $S_3$ symmetry, and have non-vanishing magnetization, given in \eqref{MPotts} and \eqref{MnotA}. Since the Hamiltonian is always invariant under the $S_3$ symmetry, we expect that this symmetry remains spontaneously broken even when the couplings are deformed away from these special points. The Potts order therefore persists in the region around $f=\lambda=0$, while the not-$A$ order remains in the region around $f=0,\,J=-\lambda$.

As the couplings are deformed away from the special points, the ground states deform, and the magnetization with them. How they transform under the $S_3$ symmetry, however, can only change if there is a phase transition. 
In finite size the degeneracy between the three ground states is split by corrections exponentially small in $L$. 
These three states are linear combinations of states  $\ket{\Psi_A}$, $\ket{\Psi_B}$, $\ket{\Psi_{C}}$ deforming those in
 \eqref{gsPotts}, still satisfying
\begin{eqnarray*}
\mathcal{S}\ket{\Psi_A}&=\ket{\Psi_B}\,,\  \mathcal{S}\ket{\Psi_B}&=\ket{\Psi_C}\,,\cr
\mathcal{C}\ket{\Psi_B} &= \ket{\Psi_C}\,,\  \mathcal{C}\ket{\Psi_A}&=\ket{\Psi_A},
\end{eqnarray*}
as the corresponding product states do. It follows using the definition \eqref{Mdef} that $M_{A}$ is real, and that 
\begin{align}
M_{B}=\omega M_{A},\qquad M_{C}=\omega^2 M_{A}\ .
\end{align}
Analogously, near the not-$A$ product state $M_{\bar{A}}$ remains real and
\begin{align}
M_{\bar{B}}=\omega M_{\bar{A}},\qquad M_{\bar{C}}=\omega^2 M_{\bar{A}}
\ .
\end{align}

By continuity, $M_g^3$ remains real and independent of $g$ while it is non-vanishing. The only way to deform between positive and negative values then requires that the magnetization must vanish at some coupling where the $S_3$ symmetry is not spontaneously broken.  We thus have an unambiguous way of distinguishing between the two phases: for any symmetry-breaking ground state in the Potts ordered phase, $M_g^3>0$, while for any symmetry-breaking ground state in the not-$A$ phase, $M_g^3<0$. The phases are separated by a phase transition with vanishing magnetization. As neither the RSPT phase nor the disordered Potts phase has local order, $M_g^3$ vanishes in these phases.

When written in terms of parafermions, the ordered Potts phase becomes a topological phase \cite{Alicea2015}. With open boundary conditions, the three ground states can be thought of as arising from the parafermionic excitations being localized at the edge. It is worth noting though that because of the presence of time-reversal and parity symmetries, this degeneracy between states only holds for the ground state.

\subsection{The RSPT phase}

The magnetization vanishes both at the Potts disordered point and at the MPS point. Distinguishing between the two phases therefore requires a subtler analysis than that needed for the ordered phases. Here we do this analysis, showing that the MPS point is part of a phase with representation symmetry-protected topological (RSPT) order.

The matrix-product ground state $\ket{\psi_{\text{MPS}}}$ in \eqref{our_MPS} is similar to the famous AKLT ground state of a spin-1 $SO(3)$-invariant chain \cite{Affleck1987, Affleck1988}. Both the corresponding Hamiltonians are special cases of a more general model with exact ground states \cite{Klumper1993,Lange1994}, whose Hamiltonian is
\begin{align*}
H_{\rm MPS} = &\sum_{j=1}^L \biggl[ h_j^2 + \beta \left( h_j g_j + g_j h_j \right)+\beta' g_j(1+g_j)\cr
&+\alpha_2 g_j^2 + \alpha_3 h_j 
+\alpha_4\left((S_j^z)^2 + (S_{j+1}^z)^2 \right) +  \alpha_0 
\biggr],
\end{align*}  
where $h_j =  S_j^+ S_{j+1}^- + S_j^- S_{j+1}^+$, $g_j  = S_j^z S_{j+1}^z$, and the couplings are related as $\alpha_0=a^2-2$, $\alpha_2 = a^2 - 2 |\beta+a|$, $\alpha_3  = a+\beta$, $\alpha_4 = |a+\beta| + 1-a^2$. For any choice of $a,\beta, \beta'$, these have a zero-energy ground state of the two-channel MPS form \eqref{our_AKLT_GS}, where 
\begin{align}
 R(j) = \begin{pmatrix}
 \ket{0}& -\sqrt{a}\ket{1} \\
 \sqrt{a}\ket{2}& -{\rm sign} (\beta + a)\ket{0}
 \end{pmatrix}\ .
 \label{MPS}
\end{align} 
Our Hamiltonian corresponds to setting $(a,\beta,\beta')=(-1,0,1/2)$, and the ensuing factors of $\pm i$ in \eqref{MPS} can be gauged away. The AKLT Hamiltonian corresponds to $(a,\beta,\beta')=(2,1,3)$.  
The most significant distinction between the AKLT ground state and ours is the minus signs. These signs, however, can be removed by the unitary transformation $H_{\rm MPS} \to \mathcal{U} H_{\rm MPS}\, \mathcal{U}^{-1}$ with $\mathcal{U} = \prod_{j} e^{i j \pi S_j^z}$, which sends $a,\beta,\beta'$ to $-a,-\beta,\beta'$. 

The AKLT model provides the canonical example of an SPT phase stable against various kinds of symmetry-preserving perturbations \cite{Gu2009,Pollmann2010}. As long as a dihedral $D_2$ symmetry of $\pi$ rotations about all three orthogonal axes $x$, $y$ and $z$ is preserved, the SPT order protects this ``Haldane phase'' as long as the gap does not close. One consequence is the four ground states in the presence of open boundary conditions persist throughout this phase. This SPT phase also is protected by either a time-reversal or parity symmetry \cite{Gu2009,Pollmann2010}. 

Owing to the similarity of the AKLT state to our MPS ground state, it is natural to expect analogous behavior in the surrounding phase.  However, none of the symmetries protecting the Haldane phase persists in our Hamiltonian. The $D_2$ symmetry comes closest. It is generated by charge conjugation $\mathcal{C}$ along with the ${\mathbb Z}_2$ symmetry $(-1)^Q$. Hence it is preserved on the $U(1)$-invariant line but not otherwise, as in general our model preserves only $\omega^Q$.  The time-reversal symmetry protecting the Haldane phase is distinct from ours and is not present here, as it is broken by the perturbation $\sigma_j^{\dag}\sigma_{j+1} + \sigma_j\sigma_{j+1}^{\dag}$. Parity symmetry is still a symmetry, but both it and our time-reversal symmetry have trivial representations for the edge states. Thus SPT order apparently survives in our model only along the $J<0$ part of the $U(1)$ line. 

The $S_3$ symmetry, however, is enough to guarantee the stability of the phase in our model.
We demonstrate the ensuing order by studying the ground states of the open-chain Hamiltonian
\begin{align}
H_{\gamma,\hat{\gamma}} \equiv \sum_{j=1}^{L-1} \left[P_{2j-1,2j}(\gamma) + P_{2j,2j+1}(\hat{\gamma})\right]\ .
\end{align}
At the MPS point with $\gamma=2$ and $\hat{\gamma}=3/2$, this Hamiltonian has four ground states given by the entries of the matrix product 
\begin{align}
\begin{pmatrix}
\ket{uu}&\ket{ud}\\
\ket{du}&\ket{dd}
\end{pmatrix} \equiv 
\left(R(1)\otimes R(2)\otimes \dots R(L)\right)\,,
\label{openGS}
\end{align}
i.e.\ \eqref{our_AKLT_GS} without the trace. 
Since the system is gapped, the edges of the system can be treated as uncorrelated. Thus one can intuit that in the ground state, each edge belongs to one of two possible states, labeled by $u$ and $d$ in \eqref{openGS}. Choosing one of the two for each edge gives the four ground states. A key observation of the work on SPT phases is that analyzing the effect of global symmetries on the ground states allows one to not only make this notion precise, but to show how such phases are robust under perturbations.

The generators of the $S_3$ group $\mathcal{C}$ and $\mathcal{S}$ obey
\begin{align}
\mathcal{C}^2=1\, ,\quad \mathcal{S}^3=1\,,\quad \mathcal{C} \mathcal{S}^2= \mathcal{S} \mathcal{C}\, ,\quad 
 \mathcal{S}^2 \mathcal{C}=\mathcal{C} \mathcal{S}\ .
\label{S3alg}
\end{align}
The action of $\mathcal{C}$ and $\mathcal{S}$ on the MPS ground states amounts to replacing the matrix \eqref{our_MPS} with 
\begin{align}
  \begin{pmatrix}
\ket{0} & \ket{-} \\
\ket{+} & \ket{0}
\end{pmatrix} ,\qquad
 \begin{pmatrix}
\ket{0}& \omega\ket{+} \\
\omega^{-1}\ket{-} & \ket{0}
\end{pmatrix} ,
\label{abcMPS}
\end{align} 
respectively.  These actions can be recast as operations in the MPS auxiliary space using
\begin{align}
{ U}_{\mathcal C} = \left( \begin{array}{cc}0&1 \\ 1 &0 \end{array}  \right)
\,,\qquad 
{ U}_{\mathcal S} = \left( \begin{array}{cc}\omega^{-1}&0 \\ 0&\omega  \end{array}  \right) \,.
\label{UCUS}
\end{align}
Then taking $R(j)\to U_{\mathcal C} R(j)U_{\mathcal C} ^\dagger$ and $R(j)\to U_{\mathcal S} R(j) U_{\mathcal S}^\dagger$ implements the symmetries on the ground states. 

When implementing the symmetry on the four MPS ground states via these unitaries, all the operations cancel except at sites $1$ and $L$. These states therefore transform under the $S_3$ via matrix multiplication at the edges, e.g.\ under a charge conjugation
\begin{align}
\begin{pmatrix}
\ket{uu}&\ket{ud}\\
\ket{du}&\ket{dd}
\end{pmatrix} 
\longrightarrow 
& \left( \begin{array}{cc}0&1 \\ 1 &0 \end{array}  \right)
 \begin{pmatrix}
\ket{uu}&\ket{ud}\\
\ket{du}&\ket{dd}
\end{pmatrix} 
\left( \begin{array}{cc}0&1 \\ 1 &0 \end{array}  \right)\cr
&=\begin{pmatrix}
\ket{uu}&\ket{du}\\
\ket{ud}&\ket{dd}
\end{pmatrix} \ .
\label{cgs}
\end{align}
The unitary operators  $U_{\mathcal{C}}$ and $U_{\mathcal S}$ obey the same relations as \eqref{S3alg}, and form an irreducible two-dimension representation of the $S_3$ symmetry group.  The same goes for their hermitian conjugates $U_{\mathcal{C}}^\dagger$ and $U_{\mathcal S}^\dagger$. Therefore {\em each} edge transforms as this two-dimensional representation of the symmetry group. In SPT states, the analogous matrices form instead projective representations of the symmetry group \cite{Chen2011,Pollmann2012}. As we explain, this distinction is why our phase is not quite as strongly protected.

The key to distinguishing the ordered phases was demonstrating that the ground states in the Potts ordered and not-$A$ phases could not be deformed into each other without closing the gap. An analogous but subtler argument applies in the disordered phases. Deforming the couplings away from the MPS state, the ground states must also transform as in \eqref{cgs} and the analog for $\mathcal{S}$. Moreover, since the Hamiltonian is local, the edges remain uncorrelated (up to exponentially small finite-size corrections). Then each edge must still transform in the doublet of $S_3$ as long as the gap does not close. The four ground states persist in a region around the MPS state as long as the $S_3$ symmetry is preserved. For any boundary conditions, the Potts disordered state has a unique ground state transforming trivially under the $S_3$. It is thus distinct from the ``RSPT'' phase in the upper left of figure \ref{phase_diagram}.

The ``R'' in RSPT is for representation, as the fact the edges transformed non-trivially under a non-Abelian representation of $S_3$ was crucial in characterizing the phase. However, the RSPT phase is not as robust as an SPT. In the preceding analysis, we considered only deformations that left the Hilbert space unchanged. Full SPT order remains robust even if other degrees of freedom are coupled in a symmetry-preserving fashion, a consequence of each edge individually transforming as a projective representation (both edges together give a conventional representation, as they must). No symmetry-preserving local perturbation can break the degeneracies in an SPT phase. However, even though our non-Abelian symmetry is strong enough to protect the order under deformations of the original chain, it is still local. Coupling the edge to an added two-state system can destroy the order, as one might expect. Namely, define
\begin{align}
\sigma^+ = \begin{pmatrix}
0 & 1 \\
0 & 0
\end{pmatrix}, \qquad \sigma^- = \begin{pmatrix}
0 & 0 \\
1 & 0
\end{pmatrix}\ ,
\end{align}
as the operators acting on the extra two-state system, which we take to transform as a doublet under the $S_3$ symmetry. Coupling it to the edge spin via
\begin{align}
H_{\text{break}} = \lambda_{\text{break}}\left(S_L^+\sigma^- + S_L^-\sigma^+\right)
\end{align}
then leaves all our symmetries intact. It is easy to check numerically that the degeneracies are split in two for $\lambda\ne 0$. 
Coupling the analogous term to the other edge then removes the edge degeneracy entirely, leaving a unique ground state. $S_3$ symmetry is no longer sufficient to protect the edge degeneracy.

The other irreducible representations of $S_3$ are one-dimensional, and their tensor product with a doublet still leaves a doublet. 
A three-state spin transforming under $S_3$ in the usual way described in section \ref{Model} is comprised of a doublet and a one-dimensional representation. The latter means that coupling an added three-state spin to the edge spin need {\em not} split the edge degeneracy. The edge-spin degeneracy is thus stable to all $S_3$ deformations preserving the Hilbert space, and to those coupling edge spins to one-dimensional representations. This is still rather robust behavior, and so we prefer the name Representation SPT (RSPT) to that of a ``fragile'' SPT sometimes used in a similar context.\footnote{We thank S. Parameswaran for suggesting the name, and the Aretha Franklin observation.}

\section{The phase transitions}

\label{Continuum_limits}

\begin{figure}[ht]
\vspace{-2.5cm}\hspace*{-0.74cm}
\includegraphics[width=1.18\linewidth]{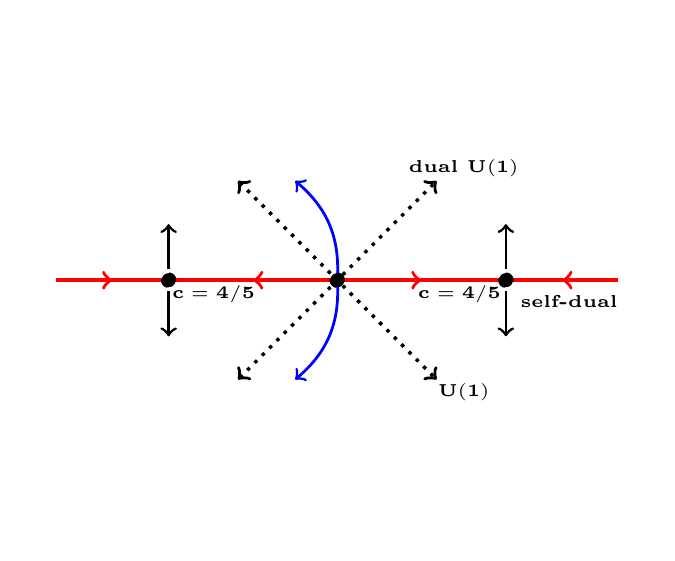}
\vspace{-3cm}\caption{RG flow about the $c=1$ (central) and $c=4/5$ points, with axes as in Fig.\ \ref{phase_diagram}. The dotted black and solid red (horizontal) lines are exact, as symmetry constrains the flow to be along the red line when self-duality is imposed and along the dotted lines for $U(1)$ or dual $U(1)$. The other critical flows starting at the $c=1$ CFT must start out vertically but no lattice symmetry constrains them to stay that way; numerics given in the section \ref{sec:numerics} indicate the curves shown.}  
\label{RG_flow}
\end{figure}

We have shown that there are at least four phases in our phase diagram in the regions surrounding each of the points with exact ground states. The phases with spontaneously broken $S_3$ symmetry are in the upper right and lower left of Figure \ref{phase_diagram}, while the RSPT order is in the upper left. The goal of this section is to understand
the phase transitions separating the four, and to rule out any other nearby phases,  

To this end, we use a field-theory analysis. The continuum limits of both $H_0$ and the self-dual Potts model $H_P(J,J)$ ($J>0$) are given by two-dimensional conformal field theories (CFTs)  \cite{Ginsparg1988a,DiFrancesco1997}.  The symmetries of the lattice Hamiltonian allow the field theories describing various perturbations to be identified precisely. Combining these observations with a known renormalization-group flow between these two conformal field theories \cite{Fateev1991,Delfino2002,Lecheminant2002}, allows us to show that the model must have four distinct phases separated by phase transitions in the three-state Potts universality class. These flows are derived in this section and summarized in Figure \ref{RG_flow}. 

Such calculations are possible because of the very strong constraints of conformal invariance in two spacetime dimensions. These constraints, along with other symmetries, are sufficient to determine all possible scaling dimensions in both the CFTs here \cite{Blote1986,Affleck1986}. Conformal transformations here can be split into holomorphic and antiholomorphic parts, acting independently on the $z$ and $\overline{z}$ complex coordinates. In the Hilbert-space (i.e.\ real-time) approach, the Hamiltonian can be split into the corresponding left and right-moving parts:
\begin{align}
H_{\text{CFT}}=\frac{2\pi}{L}\left(L_0+\overline{L}_0-\frac{c}{12}\right),
\label{HCFT}
\end{align}
The operators $L_0$ and $\bar{L}_0$ commute with each other, while the universal number $c$ is the called the central charge. States therefore can be labelled in terms of these eigenvalues as $\ket{h,\bar{h}}$, where $L_0\ket{h,\bar{h}}=h\ket{h,\bar{h}}$, $\bar{L}_0\ket{h,\bar{h}}=\bar{h}\ket{h,\bar{h}}$\,. 
An operator creating any state $\ket{h,\bar{h}}$ from the ground state has scaling dimension $h+\bar{h}$. 

\subsection{Flows near the three-state Potts CFT}

The three-state Potts CFT describing the continuum limit of $H_P$ with $f=J$ has been understood for quite some time
\cite{Dotsenko1984,Cardy1986}. 
It has central charge $c=4/5$, and a list of all the relevant operators and their symmetries can be found in Ref.\ \onlinecite{Mong2014}. 
Only one relevant operator is invariant under the $S_3$ symmetry and both parity and time-reversal symmetries. It is known as the energy operator $\epsilon(x)$, where $x$ denotes the spatial coordinate in the continuum. The energy operator has $(h,\bar{h})=(2/5,\,2/5)$, and is odd under the duality. These facts imply that the anti-self-dual lattice operator
\[
\tau_j + \tau_j^\dagger - \sigma^\dag_j\sigma_{j+1} - \sigma_j\sigma^\dag_{j+1}\quad \longrightarrow\quad \epsilon(x)
\]
in the continuum limit \cite{Dotsenko1984}.
Taking $f\ne J$ in $H_P$ corresponds to perturbing the self-dual critical point by precisely this lattice operator, indeed the reason $\epsilon(x)$ is named the energy operator. The corresponding perturbed CFT is thus
\[H_{\rm Potts} +(f-J) \int dx\, \epsilon(x)\ .\] 
This relevant perturbation of the critical point results in the vertical flows near $H_P$ in Figure \ref{RG_flow}. This perturbed CFT is integrable and gapped \cite{Zamolodchikov1988}, and no non-trivial fixed point results from this flow for either sign of $f-J$. Indeed, the Potts completely ordered and disordered Hamiltonians $H_{\rm PO}$ and $H_{\rm PD}$ treated in Section \ref{sec:Pottsexact} are gapped with correlation length zero.

Perturbing the Potts Hamiltonian by the $U(1)$-invariant self-dual operator $H_0$ is a different matter entirely. Thinking of $H_0$ as an operator in the Potts CFT, it has all the symmetries of $H_P$ plus the $U(1)$ symmetries generated by $Q$ and $\widehat{Q}$. There is no relevant self-dual operator in the three-state Potts CFT invariant under parity or time-reversal symmetry (although there is a chiral self-dual one \cite{Cardy1992}). Thus perturbing $H_P$ by $H_0$ should result in a flow back into $H_P$, as illustrated by the horizontal self-dual line near $H_P$ in Figure \ref{RG_flow}. The least irrelevant operator invariant under all the appropriate symmetries has dimensions $(7/5,\, 7/5)$, and so the flow back into  $H_P$ should be via this operator.

\subsection{The $U(1)$-invariant CFT}
\label{sec:freeboson}

Since $H_0$ has more symmetry than $H_P$, it is natural to expect that along the self-dual line $f=J$ there is a flow from $H_0$ to $H_P$. We find that this indeed is what happens for $\lambda>0$, with a beautiful continuum picture \cite{Fateev1991,Lecheminant2002}. (Such a flow, however, does {\em not} occur from $-H_0$ to $H_P$, where $\lambda$ is negative; other phases intervene \cite{OBrien2019b}.) To utilize the continuum picture, however, we must first identify the CFT describing $H_0$, and find the properties of its operators under duality and the symmetries.

Since $H_0$ has $U(1)$ symmetries, the simplest possible conformal field theory describing its continuum limit is that of a free compact boson with central charge $c=1$.  In order to make contact with the conformal field theory literature, we use a different normalization of the field as compared to the field $\Phi$ in the introduction. We define a compact bosonic field $\phi$ to take values on a circle of radius $r$, so that $\phi$ is identified with $\phi+2\pi r$. Just like the Hamiltonian \eqref{HCFT} can be split into two commuting pieces, the field can be as well:
\begin{align}
\phi(z,\bar{z}) = \varphi(z) + \overline{\varphi}(\bar z) \ ,
\end{align}
with action in Euclidean spacetime
\begin{align}S_{B}=\frac{1}{2\pi}\int d^2 z \left[(\partial\varphi)^2 + (\bar{\partial}\overline{\varphi})^2\right]\ .
\end{align}
The action and Hamiltonian are thus invariant under independent shifts $\varphi\to\varphi +b$ and $\overline{\varphi}\to\overline{\varphi} +\bar{b}$, and so has two $U(1)$ symmetries. It is customary to call the conserved charge arising from shifts in $\phi$ the {\em electric} charge, while the {\em magnetic} charge arises from shifting the dual field $\widehat{\phi}=\varphi- \overline{\varphi}$, the names stemming from the Coulomb-gas approach to critical phenomena \cite{Nienhuis1984}. 
The model has two ${\mathbb Z}_2$ symmetries given by sending $\varphi\to-\varphi$ or $\overline{\varphi}\to-\overline{\varphi}$. Doing both thus sends $\phi\to-\phi$, while doing the latter exchanges $\phi$ with $\widehat{\phi}$, and so is {\it electric-magnetic duality}.

All the possible scaling dimensions for a compact boson are contained in the partition function
\begin{align}
Z(q,\bar{q})=\text{Tr}\left(q^{L_0-c/24}\bar{q}^{\bar{L}_0-c/24}\right)\ .
\label{Zqq}
\end{align}
One can think of  $Z(q,\bar{q})$ as a partition function on a torus labeled by modular parameter $\tau$, with 
$q=\exp(2\pi i \tau)$ and $\bar{q}=\exp(-2\pi i \bar{\tau})$. This definition generalizes the usual finite-temperature partition function, as using \eqref{HCFT} gives
\[\hbox{Tr}\, e^{-H_{\rm CFT}/T}=Z\big(e^{-2\pi/(LT)},e^{-2\pi/(LT)}\big)\ .\]
For the two-dimensional free boson, $Z(q,\bar{q})$ can be computed directly \cite{Ginsparg1988a,DiFrancesco1997} for any boson radius $r$, yielding
\begin{align}
Z(r)=\frac{1}{\eta\bar{\eta}}\sum_{m,n\in \mathds{Z}} q^{\frac{1}{8r^2}\left(m+2r^2n\right)^2}\bar{q}^{\frac{1}{8r^2}\left(m-2r^2n\right)^2}
\label{ZH0}
\end{align}
with $m$ and $n$ the electric and magnetic charges respectively, while $\eta$ is the Dedekind $\eta$-function defined as
$
\eta\equiv q^{1/24}\prod_{n=1}^{\infty}(1-q^n).$
We have adopted the normalization convention \cite{Ginsparg1988a} that the operators $e^{\pm im\phi/r}$ have electric charge $\pm m$, vanishing magnetic charge, and dimensions $(m^2/8r^2,m^2/8r^2)$. Likewise, $e^{\pm 2inr\widehat{\phi}}$ have magnetic charge $\pm n$ and are of dimension $(n^2r^2/2,n^2r^2/2)$. From these expressions it is apparent that interchanging $m$ and $n$ leaves the partition function invariant when $r\to 1/(2r)$ as well.

By using the constraints coming from integrability, the precise conformal field theory corresponding to the continuum limit of  $H_0$ was identified long ago \cite{Baranowski1990}. It is indeed that of a compact boson, with radius $r=\sqrt{3/2}$. 
From the partition function \eqref{ZH0}, one can identify the left and right scaling dimensions of all operators in the theory simply by reading off powers of $q$ and $\bar{q}$ appearing in its expansion.  Setting $r=\sqrt{3/2}$ gives the scaling dimensions of all the operators appearing in the continuum limit of $H_0$ to be \begin{align}
\left(h,\bar{h}\right)=\left(\frac{(m+3n)^2}{12}+a,\frac{(m-3n)^2}{12}+\bar{a}\right),
\label{hmn}
\end{align}
where $a$ and ${\bar a}$ can be any non-negative integers.

The next task is to identify the relevant operators and their symmetry properties. Charge-conjugation symmetry can be identified with the $\phi\to-\phi$, $\widehat{\phi}\to-\widehat{\phi}$ symmetry of the CFT. Since $\mathcal{C}$ does not commute with $Q$ and  $\phi\to -\phi$ does not commute with shifts of $\phi$, the electric charge $m$ must be the eigenvalue of $Q$ in the field-theory limit. The $\mathbb{Z}_3$ charge is thus $\omega^m$, and the magnetic charge $n$ is the eigenvalue of $\widehat{Q}$. The Kramers--Wannier duality of the lattice model then becomes the electric-magnetic duality in the CFT. Indeed, under both lattice and CFT dualities a state with charges $(m,n)$ is mapped to one with charge $(n,m)$. 

In fact, the {\em only} boson radius consistent with these symmetries and the action of duality is $r=\sqrt{3/2}$. Our derivation exploits the fact that Kramers--Wannier duality on the torus mixes symmetry sectors with various boundary conditions (see e.g.\ Refs.\ \onlinecite{Baxter1982,Nienhuis1984,Schutz1993}). For example, in two-dimensional classical lattice models duality is proved by showing that the high-temperature graphical expansion is equivalent to the low-temperature expansion in terms of domain walls (oriented in our three-state case). The latter expansion only allows for certain domain-wall configurations to be wrapped around a cycle of the torus, whereas the former is not restricted, and so establishing equalities between such partition functions requires some care. One finds with a more detailed calculation \cite{Schutz1993}, that for $H_0$ with periodic boundary conditions, only the sector with trivial $\mathbb{Z}_3$ charge is invariant under duality. The other sectors instead transform to sectors with twisted boundary conditions. 

Thus the CFT partition function restricted to electric charge a multiple of $3$, i.e. $m=3m'$, is self-dual, while the full $Z(r)$ is not. We therefore require 
\[
Z^{}_{\omega^Q=1}(r)=\frac{1}{\eta\bar{\eta}}\sum_{m',n\in \mathds{Z}} q^{\frac{1}{8r^2}\left(3m'+2r^2n\right)^2}\bar{q}^{\frac{1}{8r^2}\left(3m'-2r^2n\right)^2}
\]
to be self-dual, i.e.\  unchanged by $m'\leftrightarrow n$ with $r$ fixed. This forces $2r^2=3$ and hence $r=\sqrt{3/2}$. Analogous $c=1$ points occur in models with $N$ states per site \cite{Vernier2019}, and the same duality argument can be used to show that these have boson radius $r_N=\sqrt{N/2}$.

\subsection{Flows from the $U(1)$-invariant CFT}

The $S_3$ symmetry of our Hamiltonian $H$ tightly constrains the field theory describing its continuum limit in the region around $H_0$. Since the ${\mathbb Z}_3$ part is generated by $\omega^Q$, the $U(1)$ charge modulo 3 is still preserved. Thus any perturbed CFT description of our Hamiltonian can include only operators that have $m=3m'$ for integer $m'$. Moreover, the charge-conjugation symmetry means they must also be invariant in sending $m\to -m$ and $n\to-n$.  

From \eqref{hmn} it is thus apparent that only two relevant operators are both $S_3$ and chirally invariant, both with dimensions $(3/4,\,3/4)$. The operator $\cos\sqrt{6}\phi $ violates electric charge by $\pm 3$ and preserves magnetic charge, while $\cos\sqrt{6}\,\widehat{\phi}$ preserves electric charge but violates magnetic charge. To fix the field theory precisely, note that duality corresponds to exchanging the two terms, whereas the lines $f=0$ and $J=0$ preserve $Q$ and $\widehat{Q}$ respectively. 
Ignoring all irrelevant operators, the general perturbed CFT action describing the continuum limit of $H$ is therefore
\begin{align}
S=S_{\rm B} + \Gamma\int d^2z \left[ f \cos\sqrt{6}\,\phi + J \cos\sqrt{6}\,\widehat{\phi}\right]\ ,
\label{SPCFT}
\end{align}
where $\Gamma$ is a non-universal constant symmetric in $f,J$. We thus have derived the action \eqref{SPhi} in the introduction with the rescaling $\phi= r\Phi$. 

The field theory corresponding to setting either $f=0$ or $J=0$ is a well-known one, the sine-Gordon model; for a review see e.g.\ \onlinecite{Mussardo2010}. It is integrable and gapped, so the flows along the $U(1)$-preserving lines do not reach non-trivial fixed points. In Figure \ref{phase_diagram}, these are along the lines  $\alpha=\pm \beta$. Thus all the models with exact ground states are gapped. As described in the introduction, one can compute the magnetization along the $J=0$ line directly in the field theory, in harmony with the lattice results derived above.

For $|f|\ne|J|$, the field theory remains gapped in general. However, something very special happens along the self-dual line $f=J$. It was convincingly argued \cite{Delfino2002,Lecheminant2002} that this field theory describes an integrable flow \cite{Fateev1991} from this particular free-boson field theory with $c=1$ to the Potts conformal field theory with $c=4/5$.\footnote{One small subtlety: in Ref.\ \protect{\onlinecite{Lecheminant2002}}, the $c$=1 bosonic field theory involved is called, slightly inaccurately, the ${\mathbb Z}_4$ parafermion theory. This parafermion field theory and free-boson field theory are not exactly the same, as the former is an orbifold of the latter and has no $U(1)$ symmetries \protect{\cite{Ginsparg1988b}}. Since the flow occurs in both cases, the distinction is mainly of interest when analyzing the operator content.} In the case $f=J>0$, this flow very naturally appears in our phase diagram: adding $H_P$ to $H_0$ is relevant, and causes a flow between the two critical points. Adding $H_0$ to $H_P$ is indeed irrelevant, as we have shown. 

The field theory is independent of the signs of $f$ and of $J$, because either can be flipped by redefining $\phi$ and/or $\widehat{\phi}$ by a shift of $\pi\sqrt{2/3}$. The flow is therefore the same for all $|f|=|J|$, and so occurs for all {\em four} of these perturbations of $H_0$. There are thus $c=4/5$ critical lines emanating from $H_0$ in both vertical and horizontal directions in Figure \ref{RG_flow}. 

While this field theory analysis makes these flows clear, in the lattice model only the self-dual $f=J>0$ case seems immediately apparent, in that perturbing around $H_0$ gives a relevant self-dual perturbation flowing to $H_P$. In the self-dual case $f=J<0$, the couplings in $H_P$ are antiferromagnetic, but the field theory still predicts the transition between the RSPT and not-$A$ phases is that of the ferromagnetic three-state Potts model. 

Even more striking is what happens when $f\,$=$\,-J$. While duality relates two phases on the bottom of Figures \ref{phase_diagram} and \ref{RG_flow} to their reflection on top,  the field-theory perturbation in the vertical direction is anti-self-dual. The critical phase transition in the Potts universality class still occurs as illustrated in Figure \ref{RG_flow}, but since no lattice symmetry protects the location of these lines, they need not stay vertical. The numerics discussed below in section \ref{sec:numerics} are needed to locate the transition precisely.  Finding a continuous transition from the not-$A$ phase to the disordered phase perhaps is not so surprising. However, the transition from the ordered Potts phase with spontaneous symmetry breaking directly into the RSPT phase is much more unusual -- the local order parameter of the former is quite different from the non-local order parameter of the latter. 

\section{The full phase diagram}
\label{sec:numerics}

We have found four distinct phases of our Hamiltonian $H$ in the region of $H_0$. We also have shown by the perturbed CFT analysis that four critical lines in the three-state Potts universality class terminate at the multicritical point $H_0$, and that these lines separate the different phases, as indicated in the flow diagram in Figure \ref{RG_flow}. The simplest and most natural way of putting this information together is in the phase diagram displayed in Figure \ref{phase_diagram}. 

In this section we present numerics strongly supporting this picture, and indicating that there are no other phases in this region. We also locate the vertical critical lines describing the ordered/RSPT and disorder/not-$A$ transitions.
We use the Density Matrix Renormalisation Group (DMRG) with ITensor \cite{ITensor}. For the order parameter, a lattice length of $L=200$ and bond dimension $\chi=300$ is used, while for the ground states a bond dimension $\chi=800$ is chosen.  

We determine the location of the phase transition by computing the bipartite entanglement entropy of a periodic chain using the DMRG. For a periodic system of length $L$, the leading contribution to the entanglement entropy of a CFT scales as $S(L)=\frac{c}{3}\log(L)+\text{const}$, where $c$ is the central charge of the CFT \cite{Calabrese2004}. For a gapped system, the entanglement entropy tends to a constant by the area law \cite{Eisert2010,Hastings2007}, and so the coefficient of the log term vanishes. This computation thus both allows us to find critical points and characterize them
precisely. 

\begin{figure}[ht]
\vspace{-2.8cm} \hspace*{-0.9cm}
\includegraphics[width=1.2\linewidth]{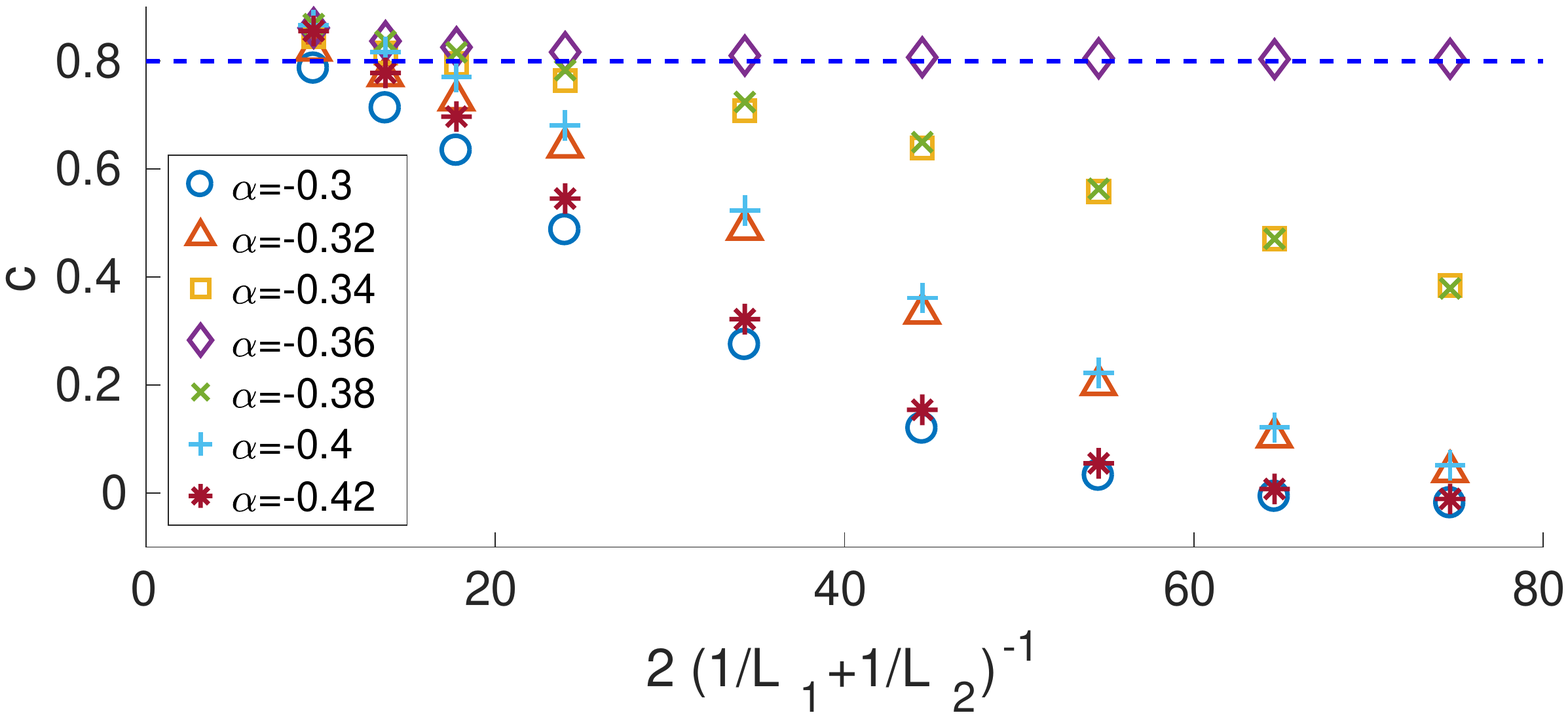}
\vspace{-1cm}\caption{The effective central charge $c$ at $\beta=0.75$ from the DMRG for $-0.3\geq \alpha\geq -0.42$. We extract it from \eqref{ceff} for consecutive $L_1,L_2$ in the list 8,12,16,20,30,40,50,60,70,80.
For $\alpha\approx -0.36$ the data show $c_{\text{eff}}\to 0.8$, while for smaller and larger values it decays to 0 with increasing $L$.   }  
\label{beta_075}
\end{figure}

By finding the effective central charge at different $\alpha,\beta$, we can locate the transition. We extract effective central charges for each $\alpha$ at a given $\beta$ using 
\begin{equation}
c_{\text{eff}}=3\frac{S(L_2)-S(L_1)}{\log(L_2/L_1)}\ .
\label{ceff}
\end{equation}
We give an example for $\beta=0.75$ in Figure \ref{beta_075}, plotting $c_{\rm eff}$ against $2(1/L_1+1/L_2)^{-1}$.
The phase transition is clearly apparent here for $\alpha\approx -0.36$, and the central charge $c=4/5$ at the transition predicted by the field theory is confirmed. We plot various transition values by the green crosses in Figure \ref{phase_diagram}. We note that the field theory argument indicates that the transition line should be vertical right at the origin, but since the numerics become rather difficult near the multicritical point, we were unable to confirm this prediction. 

\begin{figure}[ht]
\vspace{-2.8cm} \hspace*{-0.9cm}
\includegraphics[width=1.2\linewidth]{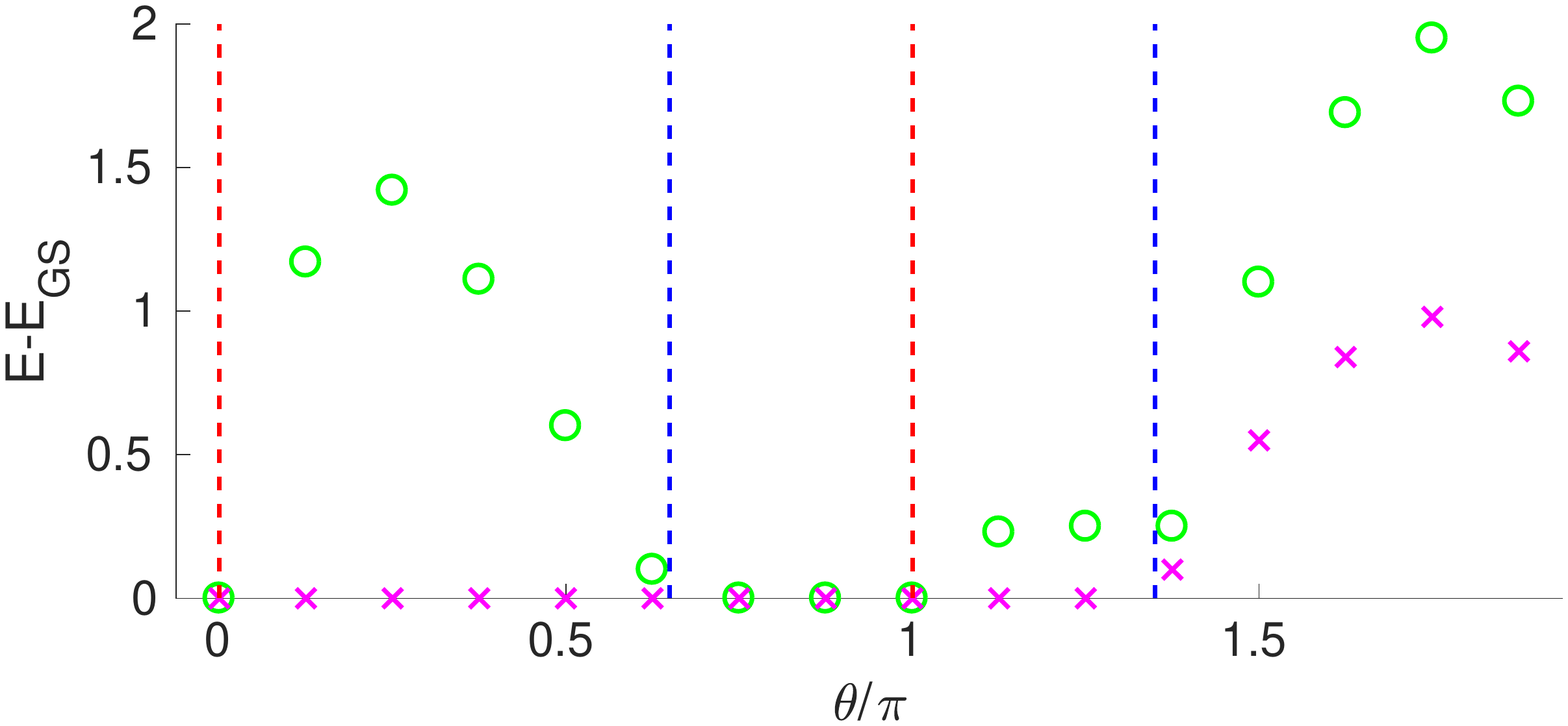}
\vspace{-1cm}\caption{The energy gaps to the lowest-energy state in the $s=1$ sector ($E_1^0-E_0^0$, magenta crosses) and to the first excited state in the $s=0$ sector ($E_0^{1}-E_0^0$, green circles) for open boundary conditions from DMRG, as a function of coupling, where $\alpha=\cos\theta$, $\beta=\sin\theta$. The positions of the phase transitions on the self-dual and numerically determined lines  are given by the red and blue dashed lines respectively.}  
\label{E_vs_theta}
\end{figure}

We analyze several properties of the phases themselves by tuning the couplings along the circle $\alpha^2+\beta^2=1=(f^2+J^2)/2$, so as to go through all the phases. In Figure \ref{E_vs_theta} two different gaps are plotted as a function of $\theta$, defined by $\alpha=\cos\theta$, $\beta=\sin\theta$. We label energies by $E_s^k$, where $s$ is the $\mathbb{Z}_3$ charge $\omega^s$ of that state, while the superscript $k=0,1,\dots$ labels which the states in that sector in order of increasing energy. In the figure we plot both $E_1^0-E_0^0$ and $E_0^{1}-E_0^0$. We take open boundary conditions so that we can observe the multiple ground states in the phase around the MPS ground state.

The region between the first two vertical dashed lines is the ordered Potts phase, where $E_1^0$=0 but $E_0^1>0$ and hence there are three degenerate ground states (charge-conjugation symmetry means that the spectra in the $s=-1$ and $s=1$ sectors are identical). Between the transitions denoted by the dashed lines at $\theta/\pi\approx 0.65$ and $\theta=\pi$, we find both gaps vanishing, up to exponentially small corrections. Thus the data display the four-fold degeneracy throughout the phase, as predicted by our RSPT analysis. The not-$A$ phase is like the Potts ordered phase, with a three-fold degeneracy among ground states, as the symmetry analysis predicts. Finally, there is a lone ground state throughout the disordered phase. Moreover, we find the spectrum above the ground state(s) is clearly gapped away from the phase transitions.

\begin{figure}[ht]
\vspace{-2.8cm} \hspace*{-0.9cm}
\includegraphics[width=1.2\linewidth]{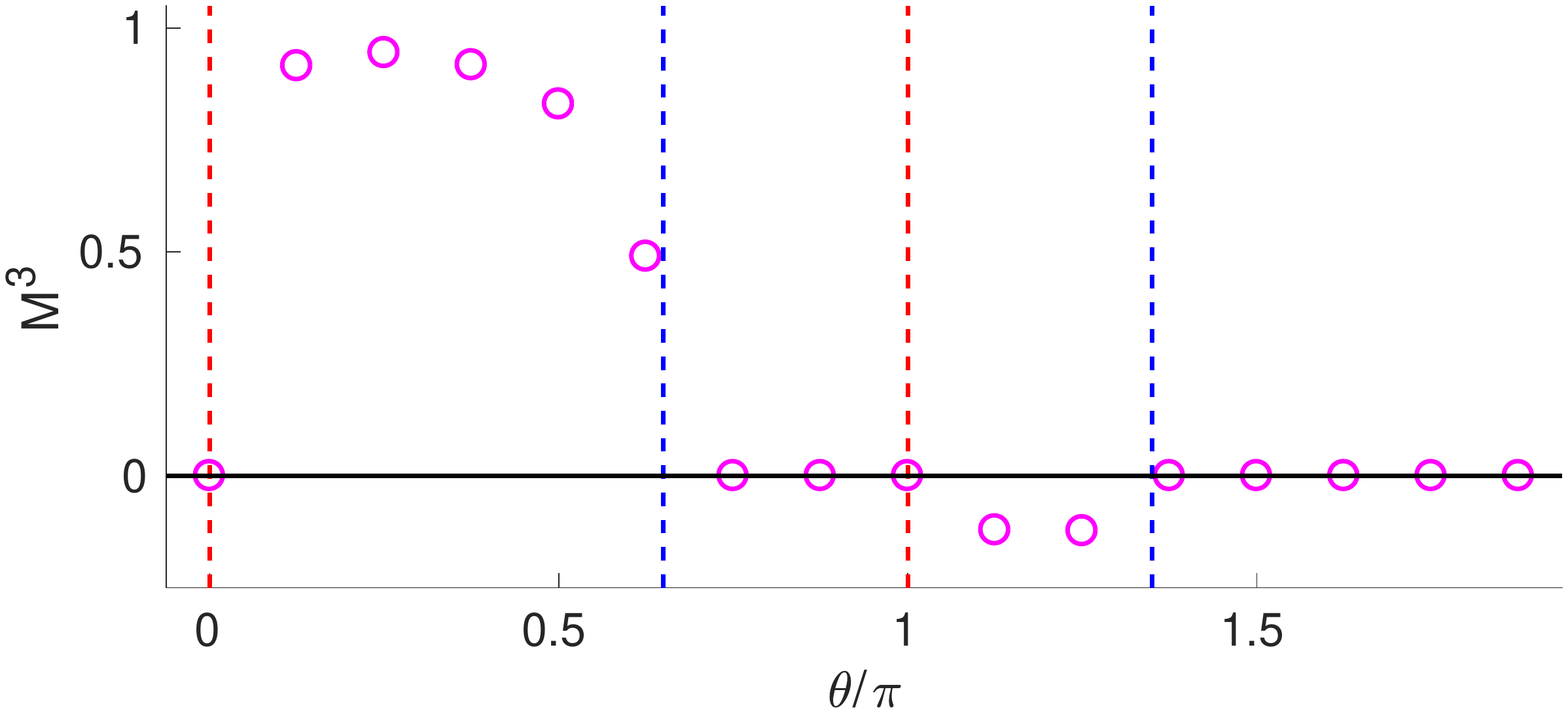}
\vspace{-1cm}\caption{The order parameter $M^3$, plotted vs.\ $\theta$ as in Fig.\ \ref{E_vs_theta}.}
\label{G_inf_vs_theta}
\end{figure}

In Figure \ref{G_inf_vs_theta} we plot the magnetization $M^3$ defined by
\begin{align}
M^3&=  \lim_{|j-k|,|k-l|,|j-l| \to\infty} G_{jkl};\cr
G_{jkl}&\equiv \langle g| \sigma_j \sigma_k \sigma_l|g\rangle\ .
\label{M3def}
\end{align}
This definition coincides with the earlier definition up to finite-size effects, with the advantage that $G_{jkl}$ is $S_3$ invariant, and so independent of ground state $\ket{g}$. 
Our DMRG numerics find that it is indeed non-vanishing with the predicted sign throughout the Potts ordered and not-$A$ phases, and vanishes elsewhere.

\section{Conclusions}
We have studied a nearest-neighbor $S_3$-invariant spin chain, and found four distinct gapped phases meeting at a multicritical point. The phase diagram is given in Figure \ref{phase_diagram}. Whereas two of these phases are the well-studied ordered/topological and disordered phases of the three-state Potts model, two of them do not seem to have been analyzed in detail before. One is an RSPT phase protected by the non-abelian $S_3$ symmetry. The ground state at a special point in this phase is a matrix-product state similar to that of AKLT, but slightly simpler: it is an equal-amplitude sum, with no factors of $\sqrt{2}$ and no minus signs. More strikingly, it behaves very nicely under duality, transforming to a product state we dubbed the not-$A$ state, with a Hamiltonian that remains nearest-neighbor. The corresponding not-$A$ order spontaneously breaks the $S_3$ symmetry, with each ground state favoring two of the three spin directions. The two ordered phases can be distinguished by positive and negative values of the local order parameter, the magnetization cubed.

By an RG analysis, supported by numerical checks, we showed that the phase transition lines are all of the critical three-state Potts universality class, including a transition from the RSPT phase to the Potts ordered phase. Our model thus gives four distinct lattice realizations of the flow between free-boson and $c=4/5$ conformal field theories \cite{Fateev1991,Lecheminant2002}. The two along the self-dual line require no further tuning, as opposed to a two-dimensional lattice model exhibiting this flow \cite{Delfino2002,Otsuka2004}.

The special point \eqref{MPS_H} in the RSPT phase with an exact MPS ground state has another remarkable feature. It possesses exact excited states \cite{Moudgalya2020} as does its cousin, the AKLT chain \cite{Arovas1989,Moudgalya2018a,Moudgalya2018b}. The findings include a hierarchy of such states that do not seem to have an analog in AKLT. Moreover, the duality yields a few exact excited states at the not-$A$ completely ordered point as well.

The interesting phases of our Hamiltonian are not exhausted by the four studied here. In a companion paper \cite{OBrien2019b}, we analyze this model along the self-dual line in different parameter regimes. Our findings include a tricritical point generalizing that found for two-state system \cite{Rahmani2015b,OBrien2018}, a self-dual gapped phase, and an unusual critical but not conformally invariant phase. We find it remarkable that such rich structure occurs in a nearest-neighbor three-state model.

\nocite{Ginsparg1988b}

\begin{acknowledgments}
We are very grateful to Sid Parameswaran and Ashvin Vishwanath for being fonts of wisdom on SPTs, in particular enlightening us to the concept (and the name) of RSPTs. This work was supported by EPSRC through grant EP/N509711/1 1734484 (EOB) along with grants  EP/S020527/1  and EP/N01930X (EV and PF).
\end{acknowledgments}

\bibliography{References_v2}

\end{document}